\documentclass[11pt]{article}

\usepackage[english]{babel}
\usepackage[margin=2.5cm]{geometry}
\usepackage{setspace}
\usepackage{graphicx}
\usepackage{color}
\usepackage[breaklinks]{hyperref}
\usepackage[defaultlines=4,all]{nowidow}
\usepackage{tabularx}
\usepackage{soul}
\usepackage{amsmath}
\usepackage{dcolumn}
\usepackage{array}
\usepackage{csquotes}
\usepackage{ragged2e}

\usepackage{makecell}
\usepackage{dcolumn}
\usepackage{floatrow}
\newfloatcommand{capbtabbox}{table}[][\FBwidth]
\usepackage{blindtext}

\usepackage[backend=biber,citestyle=authoryear,isbn=false,url=false,eprint=false,sorting=none,uniquename=init]{biblatex}
\title{Spatially concentrated social capital of urban residents}
\author{Ádám J. Kovács$^{1,*}$, Sándor Juhász$^{1,2}$, Eszter Bokányi$^{1,2}$, Balázs Lengyel$^{1,2}$}

\date{{\footnotesize%
    $^1$ELKH Centre for Economic and Regional Studies, Agglomeration and Social Networks Lendület Research Group\\%
    Budapest, H-1097, Hungary\\%
    $^2$Corvinus University of Budapest; Laboratory for Networks, Technology and Innovation\\%
    Budapest, H-1093, Hungary\\%
    $^*$Corresponding author: \href{mailto:kovacs.adam@krtk.hu}{kovacs.adam@krtk.hu}}\\[2ex]%
}

\begin{document}

\maketitle

\onehalfspacing

\begin{abstract}
Social connections that span across diverse urban neighborhoods can help individual prosperity by mobilizing social capital in cities. Yet, how the detailed spatial structure of social capital varies in lower- and higher-income urban neighborhoods is less understood. This paper demonstrates that the social capital measured on social networks is spatially more concentrated for residents of lower-income neighborhoods than for residents of higher-income neighborhoods. We map the micro-geography of individual online social connections in the 50 largest metropolitan areas of the US using a large-scale geolocalized Twitter dataset. We then analyze the spatial dimension of individual social capital by the share of friends, closure, and share of supported ties within circles of short distance radiuses (1, 5, and 10~km) around users' home location. We compare residents from below-median income neighborhoods with above-median income neighborhoods, and find that users living in relatively poorer neighborhoods have a significantly higher share of connections in close proximity. Moreover, their network is more cohesive and supported within a short distance from their home. These patterns prevail across the 50 largest US metropolitan areas with only a few exceptions. The found disparities in the micro-geographic concentration of social capital can feed segregation and income inequality within cities harming social circles of low-income residents.
\end{abstract}

\maketitle

\section{Introduction}

Social networks are essential channels to share information, knowledge, and opportunities. Connections can help people in getting a job (\cite{granovetter1995}), in achieving higher wealth (\cite{eagle2010network}) or career progress (\cite{seibert2001}). One possibility to theorize processes behind these widely observed phenomena is the concept of social capital that stresses the importance of resources that individuals can mobilize through their social connections (\cite{coleman1988social, putnam2000, lin2001}). The structure of social networks is key in characterizing social capital (\cite{borgatti1998measures}). For example, network cohesion - often measured by triadic closure - is an important building block of social capital as it facilitates trust and the emergence of norms, while decreasing misbehavior (\cite{coleman1990book, granovetter2005}). In economics, common partners are thought to support the sustainability of cooperation links; thus, such supported ties are claimed essential for the accumulation of social capital (\cite{Jackson2012socialcap}).

An early review of \cite{mohan2002placing} stresses that geography is also an important factor in shaping social 
capital. For example, increasing geographical distance decreases both the probability of social ties (\cite{liben2005geographic, Lengyel2015}) and the probability of closed triads (\cite{Lambiotte2008}), implying the existence of spatial limits of social capital. \cite{bathelt2004clusters} and \cite{gluckler2007economic} discuss the importance of geographically non-local ties that are thought to boost economic prosperity by providing access to new knowledge and opportunities. Applying a weighted network approach, evidence suggests that individuals with spatially diverse social networks are wealthier than individuals whose connections concentrate in certain locations (\cite{eagle2010network}). Yet, this previous research  does not consider structural measures of social capital. Thus, how triadic closure and supported ties at geographical distance are related to the economic prosperity of individuals is still unknown.

In this paper, we aim to establish the empirical link between the structural measures of the social capital of spatially projected social networks and income characteristics of home neighborhoods. This is done by using large-scale social media data that enables us to investigate individual social networks across a wide range of cities. We demonstrate that neighborhood-level income is related to the spatial structure of social capital in the vast majority of investigated cities. In particular, we show that social capital of residents in low-income neighborhoods are spatially more concentrated than that of residents living in high-income neighborhoods.

Cities have been long looked at as major places of the accumulation of social capital (for a comparison between individual and collectivist social capital in cities, see \cite{portes2000two}). \cite{jacobs1961book} has stressed the spatial dimension of social capital very early by arguing that social ties of residents concentrate around their home neighborhood. This early conjecture has recently been confirmed with Facebook data from New York City: on average, around 40\% of users' friends live within 10 miles from their homes (\cite{Bailey2020}). Paired with the prevailing assortative mixing of individuals in cities (\cite{Morales2019, Dong2019, Wang2018}), this massive spatial concentration of social capital in cities has far-reaching consequences on neighborhood dynamics and can be associated with segregation and inequalities. In those cities where income difference is a major factor of assortative mixing such that social capital of the rich and poor are hardly overlapping, income inequalities rise (\cite{Tothetal2021inequality}). Related case studies from Bangladesh (\cite{bashar2019social}) and Hungary (\cite{berki2017}) show that the concentration of social capital in poor urban neighborhoods facilitates trust and stronger social norms, but at the same time, limits social mobility. Such patterns can also be observed in large-scale data: the spatial concentration of Facebook friendships in New York City is stronger in areas with a lower average income and a lower level of education (\cite{Bailey2020}).

We contribute to the above discussion by comparing the spatial dimension of individual social capital - captured by structural measures - in relatively poorer versus relatively richer neighborhoods across the 50 largest metropolitan areas in the United States. This is done with a large Twitter dataset on geotagged tweets from which the home location of users, their mutual follower network, and the home location of their connections can be identified. 
We project the individual level ego-networks of users on census tracts that enables us to infer the average household income of the ego and alters by census tract level information from the American Community Survey. This way, we can map the structure and geography of social ties for more than 80,000 users across 50 US cities. Finally, we measure three aspects of individual social capital concentration in urban space within a circle of short distance radius (1, 5, and 10~km) around the home location: the share of connections within the circle, triadic closure within the circle, and the share of supported ties within the circle.

Findings confirm that users living in neighborhoods of below median household income have spatially more concentrated social networks. Residents of lower income neighborhoods are embedded in social networks that are significantly more cohesive and include significantly more supported ties within a close proximity. These patterns prevail, with few exceptions, in the 50 largest metropolitan areas of the United States. A significant negative correlation between the continuous value of neighborhood income and spatial concentration of social capital give further support to the finding. Furthermore, spatial concentration of social capital is associated with income assortativity, because most spatially close connections of the poor are residents of disadvantaged neighborhoods, among whom triadic closure is also more prevalent.

The study highlights that the social network of people from lower income neighborhoods can offer limited access to opportunities in cities. We provide novel insights into individual level social capital measures by mapping the concentration of triadic closure and supported ties in urban social networks. Moreover, we present individual level social network features that can feed the segregation and inequality patterns observed inside cities.

\section{Data and technical approach}

We use a unique database obtained from the online social network site Twitter. A fraction of tweets (i.e. short user messages on the site) are ‘geotagged’ meaning that they have meta-information on the location from which they were sent. These tweets originate from users who enabled the exact geolocation option on their smartphones. Our dataset contains tweets collected between 2012 and 2013, and due to careful sample selection described in \cite{Dobos2013}, this dataset enables the study of spatial patterns of human behavior, see (\cite{Bokanyi2016, Kallus2015, Kallus2017, bokanyi2017prediction, Bokanyietal2021}).

\begin{figure}[hbt!]
\centering
\includegraphics[width=\linewidth]{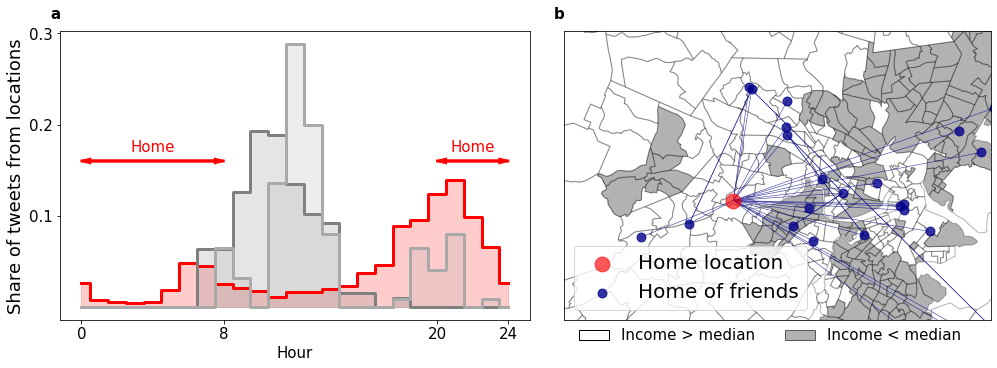}
\justify
\footnotesize{\textbf{Figure 1} Identification of home location and the spatial projection of individual social networks. \textbf{(a)} Home locations are identified as the places with the highest share of tweets in the morning (0:00-8:00) and in the evening hours (20:00-24:00). The histograms illustrate the process through an example user. The selected home cluster is marked by red. \textbf{(b)} Spatially projected social network of an example user. The home of ego is marked with red and home of alters are marked with blue. Home locations are characterized by the relation of the average household income in the respective census tract to the median income in the city.}
\end{figure}

To identify the home location of users, the Friend-of-Friend algorithm (\cite{huchra1982groups}) was used to cluster messages in space. Any two tweet coordinates are considered to belong to the same cluster, if their separation is less than 1~km. For each cluster, the first two moments of the coordinate distribution are determined. Before calculating the mean coordinates of the cluster, data points are trimmed until all points are inside a 3$\sigma$ radius to eliminate outliers. For more details about the process, see (\cite{Dobos2013, Kallus2017}). We focus on the three highest cardinality clusters with minimum 50 tweets on weekdays. The cluster with the highest share of tweets sent between 8PM and 8AM on weekdays is identified as the home location of users. Figure 1a illustrates this process through an example user.

The online social network of users is defined by their mutual followership ties on the site. To concentrate on meaningful ego networks we only consider users with at least 10 geolocated friends in the same metropolitan area. The spatially projected social network of an example user is visualized in Figure~1b. It is important to note that ties ranging outside the focal metropolitan area are disregarded because this allows a clearer comparison of social capital around home location of the rich and the poor. Additionally, we use American Community Survey (2012) data on average household income and population at census tract level to proxy the socio-economic characteristics of home locations. For simplification, we calculate the median income of census tracts for each metropolitan area and categorize each tract as above- or below the median income. Our final sample consist of 86,177 users. The distribution of users across the 50 metropolitan areas are available on Figure~S1 in the Supplementary Material.

We illustrate the spatial concentration of social ties with three different measures using concentric circles around the home locations. First, we calculate the share of friends (number of friends divided by the total number of friends) within these circles cumulatively, similarly to (\cite{Bailey2020}). Second, we also consider the change in this measure by taking its first derivative. Third, we count the number of additional friends within each subsequent concentric annulus normalized by their area to obtain spatial density. Notably, we exclude all ties of users that are farther away than 10~km distance and all measures are then calculated on subnetworks of users.

We also consider two approaches to quantify the structural cohesion of connections close to home. We first measure the local clustering coefficient of ego users in their subnetworks within concentric circles around their identified home locations. The formula of this metric is given by:

\begin{equation}
Clustering_r=\frac{2L_i}{k_i(k_i-1)},
\end{equation}
where $L_i$ stands for the number of links between the $k_i$ neighbors of node $i$ within given radius $r$. This allows us to capture the degree to which the neighbors of a given node link to each other. Though neighborhood is meant in terms of social distance, these nodes represent people who actually live in the close neighborhood in physical space as well.  

Second, we use the supported tie measure related to individual social capital introduced by (\cite{Jackson2012socialcap}). Contrary to local clustering that is calculated for nodes, support is an edge characteristic. A social tie is considered to be supported, if the two nodes linked by the relationship have at least one common partner. The formula for the share of supported ties in the subnetwork of users is the following:

\begin{equation}
Tie\, support_r=\frac{|{j\in N_i(g):[g^2]_{ij}>0}|}{d_r},
\end{equation}

\noindent where the numerator is the absolute count of the number of friends $j$ of $i$’s network $g$, who are supported by at least one friend in common within radius $r$, and $d_r$ is the degree of the node (total number of friends) within radius $r$.

\begin{figure*}
  \label{tab:table1} 
\begin{minipage}{0.38\textwidth}
\centering
\includegraphics[width=0.9\linewidth]{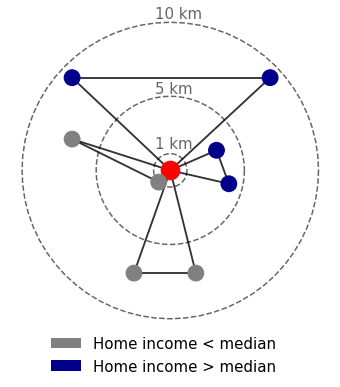}
\end{minipage}
\begin{minipage}{0.58\textwidth}
\begin{tabular}{lccc} \hline
          & \makecell{Full \\graph} & \makecell{Income \\ $<$ median} & \makecell{Income\\ $>$ median} \\ \hline
          Degree in 10 km & 8 & 4 & 4 \\ \hline
          Share of ties in 1 km  & 0.125 & 0.25 & 0 \\
          Share of ties in 5 km  & 0.375 & 0.25 & 0.5 \\
          Share of ties in 10 km  & 1 & 1 & 1\\ \hline
          Clustering in 1 km & - & - & -\\
          Clustering in 5 km & 0.333 & - & 1 \\
          Clustering in 10 km & 0.143 & 0.6 & 0.6 \\ \hline
          Support in 1 km & - & - & - \\
          Support in 5 km & 0.667 & - & 1 \\
          Support in 10 km & 1 & 1 & 1 \\ \hline
        \end{tabular}
  \end{minipage}
\justify
\footnotesize{\textbf{Table 1} Illustration of our key variables on an example graph. The red node represents our focal user. The color of the nodes refers to the income level at the census tract of users. The table describes the social capital related measures calculated within 1-5-10~km for the focal user on the example graph.
} 
\end{figure*}

Table~1 illustrates our key measures with the help of an example graph. To construct these variables we take the subnetwork of each users' ego network by different distance thresholds. The focal user has a total degree of 8 within a 10~kilometer distance from the identified home location. 37.5\% of the user's friends live within 5~kilometers distance and one of the friends (12.5\%) within 1~kilometer. The example user has 3 friends inside 5~kilometers, 2 of whom follow each other. Therefore, the local clustering coefficient inside 5~kilometers is 0.333, while in 10~kilometers, it is 0.143.
Within 10~kilometers, the share of supported ties is 1 as all of them are supported by at least one friend. 66.7\% of the ties in the example graph are supported within 5~kilometers.

To consider assortative tie formation, we construct all these variables on income-based subnetworks of users. More precisely, we create two more ego networks for each individual, where we only consider friends living in above median or below median income level neighborhoods. This allows us to measure whether these structural features are more prominent inside income groups. Table 1 illustrates that the example user has 25\% of ties in only 1~km towards lower income friends, but the network is more clustered and supported amongst wealthier people in 5~kilometers.

\section{Results}

To illustrate the concentration of individual social capital inside cities and to compare low- and high-income neighborhoods in this respect, we use three approaches in Figure~2 that quantify concentration of social connections within concentric circles around home locations. As Figure~2a illustrates, an average user living in either a below- or above-median income neighborhood has between 40\% and 50\% of their ties within a 10~kilometers radius around the home location. We find that concentration of ties is stronger for people living in lower income neighborhoods (home income $<$ median in the respective city) than for people living in higher income areas. This difference is more prevalent in Figure~2b, where we plot share of ties binned into distance categories from home that enables us to evaluate typical distances of concentration. Residents of lower-income neighborhoods are found to have the highest share of ties who live 2.5~km away while residents of higher-income neighborhoods have most friends 4.5~km away. In Figure 2a, the average cumulative share of ties decreases sharply as we focus on small radius circles around home that is due to the fact that smaller circles include lower number of residents. Therefore, we measure the density of ties within circles of distance $r$ around home (measurement is described in Data and Technical Approach section) and find a monotonously decreasing probability of ties as distance grows in Figure 2c. This monotonous decrease is less steep for lower-income areas than for higher-income ones in the distance regime where Figure~2b also shows the separation of the two income classes. This shows that there is a general trend towards the spatial concentration individual social networks in micro-space.

\begin{figure}[hbt!]
\centering
\includegraphics[width=\linewidth]{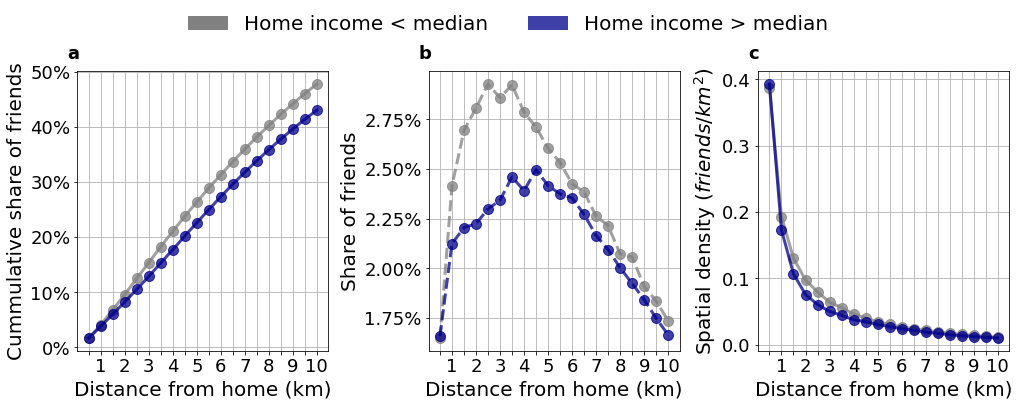}
\justify
\footnotesize{\textbf{Figure 2} Concentration of social ties around the home location of users in the top 50 metropolitan areas of the US. \textbf{(a)} Cumulative share of connections in 10~km distance from home location. \textbf{(b)} Average share of connection at the distances from home in 10~km. \textbf{(c)} Probability density of ties in 10~km. (Normalized by the respective areas). All three figures represent average values for users across the top 50 metropolitan areas of the US. We only consider users with at least 10 connections with identified home location. Median income is calculated for each metropolitan area separately.}
\label{fig:fig2}
\end{figure}

\begin{figure}[hbt!]
\centering
\includegraphics[width=\linewidth]{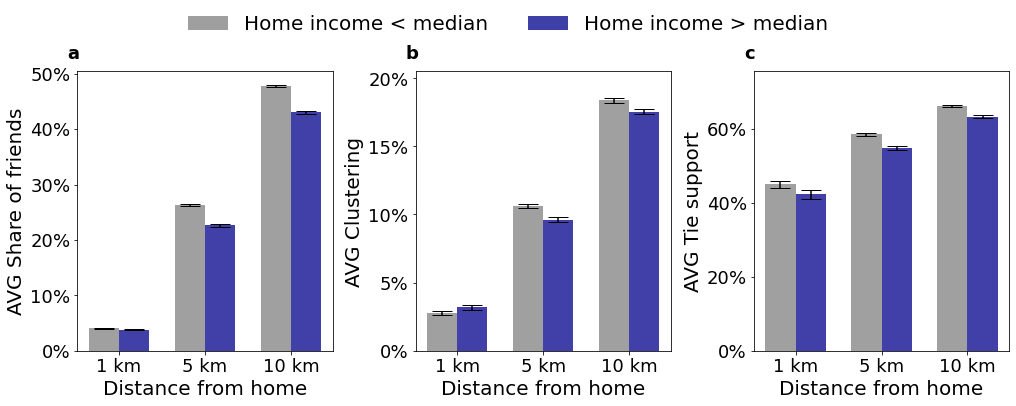}
\justify
\footnotesize{\textbf{Figure 3} Structural features of individual social networks inside 10~km. \textbf{(a)} Average cumulative share of connections in 1/5/10~km for above and below median income user groups. \textbf{(b}} Average local clustering coefficient for higher and lower income groups in 1/5/10~km. \textbf{(c)} Average tie support for users in above or below median income neighborhoods in 1/5/10~km from their home. Error bars represent 95$\%$ confidence intervals from bootstrap sampling. 
\label{fig:fig3}
\end{figure}

In Figure 3, we proceed by comparing the $95 \%$ confidence intervals of social capital measures of lower- and higher-income neighborhood residents. In general, there is a clear pattern that users from lower-income neighborhoods have a spatially more concentrated ego-network, and on average, their ties are more cohesive and involve a higher share of supported relationships in close proximity to their home. 
Figure 3a illustrates that the share of ties is significantly higher in lower-income neighborhoods in 1, 5 and 10~km circles. Measuring the average of local clustering in the ego-networks of individuals trimmed to concentric circles around their home tells us in Figure 3b that higher-income neighborhoods concentrate significantly more closed triads in 1~km than lower-income neighborhoods. Because triadic closure is negatively correlated with degree, the lower-income neighborhoods having more connections within the 1\~ km circle might lead to this difference. Therefore, it is even more striking that triadic closure is significantly higher in the 5 and 10~km circles for lower-income neighborhoods despite a higher share of friends for the same distance thresholds, with more than 10\% of the triads being closed. Concerning the share of supported ties, Figure 3c documents that spatial concentration of supported ties is the highest among the social capital indicators. More than 40\% of supported ties are within 1~km and more than 60\% within 10 ~ km. Residents of lower-income neighborhoods have significantly higher shares of supported ties in all distance categories.

In Figure 4a-c, we illustrate the relationship between the income category of home neighborhood and the spatial concentration of social capital in each of the 50 metropolitan areas. The precise description of this controlled correlation can be found in Section S3 of the Supplementary Material. Positive coefficients mean that users in the respective city with above median income tend to have higher share of their friends, higher triadic closure, and higher share of supported ties in 10~km from their home. Results suggest that in most of the cities residents of below-median income neighborhoods have higher share of connections (Figure 4a) and higher number of supported ties (Figure 4c) in 10~kilometer distance. In case of the local clustering coefficient (Figure 4b), the picture is more mixed (19/50 cases the coefficient is positive) but exceptions from the general trend are only few in case of share of friends (7/50) and the share of supported ties (12/50). Interesting exceptions are San Francisco, Detroit, Baltimore, and New Orleans, which suggests that both prosperous and segregated cities can deviate from the trend. However, the general tendency is clear that people living in lower income neighborhoods have more concentrated social capital. Strong relationships along all three dimensions are observed for both metropolitan areas with high population such as Los Angeles and Miami, and at metro areas with smaller population such as Buffalo, Raleigh and Salt Lake City.

\begin{figure}[hbt!]
\centering
\includegraphics[width=0.9\linewidth]{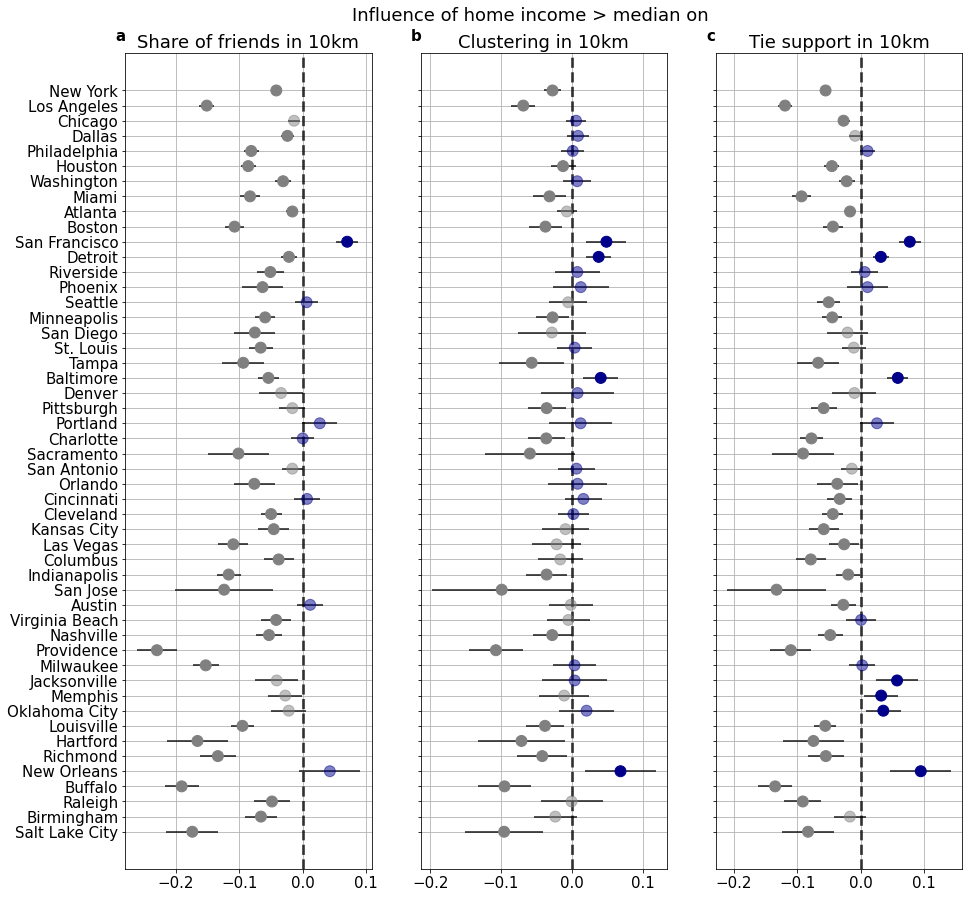}
\justify
\footnotesize{\textbf{Figure 4} Controlled correlation between above median home income and social capital related measures in 10~km from home location. Dots represent linear regression coefficients for each metro areas (see Section S3 of the Supplementary Material for more details on the models). The color of nodes indicate the sign of coefficients (gray ) and horizontal lines represent 95$\%$ confidence intervals. Nodes with lighter colors indicate statistically insignificant coefficients.} 
\label{fig:fig4}
\end{figure}

To further increase the robustness of these findings, Table~2 presents linear multivariate regression models, in which we test the correlation between the continuous value of neighborhood average income and social capital concentration within concentric circles around the home location. The table contains models in which the dependent variable is social capital within 10~km and further models including social capital concentration in 1 and 5~km are presented in Table~S1 in the Supplementary Material. In all regressions, the respective metropolitan areas are used as fixed effects that allows for sorting out the role of income differences across cities. 
We find that home income level has a statistically significant negative association with the spatial concentration of social capital. This result strengthens our claim that lower-income neighborhoods tend to have spatially more concentrated social capital than higher-income neighborhoods. 

We include the population of the census tract as a control variable because these are not uniform even within metropolitan areas. As expected, population is significantly associated with the social capital measures. The negative correlation between population and share of friends within 10~km implies that people in populous neighborhoods tend to be connected to others in distant neighborhoods. On the contrary, the positive correlations between population and triadic closure and share of supported ties suggest that the size of neighborhoods facilitates spatial concentration of social capital by enabling network cohesion. (Population is negatively associated with triadic closure and has no significant relationship with supported ties in the 1~km circle, as reported in Supplementary Material.)

Moreover, we also include the overall degree of users (irrespective of the distance of friends) as a further control variable. This way, we make sure that our results on the spatial concentration of the social network measures are not confounded by some users simply having an excess number of friends compared to others. The negative correlation between degree and share of friends within 10 km distance implies that having more connections goes together with a more widespread social network. A similar negative relationship is found for the clustering coefficient, which is unsurprising due to more friends having a lower probability of knowing each other. Conversely, the share of supported ties correlates positively with degree, implying that having more connections increases the probability of having at least one common partner with any one friend.

Further generalized robustness checks of the results in the form of regressions with dependent variables being the same measures within 1 and 5 km distance from the home locations of users can be found in Supplementary Material Section S2.

\begin{table}[hbt!] \centering 
  \label{tab:table2} 
\small 
\begin{tabular}{@{\extracolsep{10pt}}lD{.}{.}{-3} D{.}{.}{-3} D{.}{.}{-3} } 
\\[-1.8ex]\hline 
\hline \\[-1.8ex] 
 & \multicolumn{3}{c}{In 10 km from home location} \\ 
\cline{2-4} 
\\[-1.8ex] & \multicolumn{1}{c}{Share of friends} & \multicolumn{1}{c}{Clustering} & \multicolumn{1}{c}{Tie support} \\ 
\\[-1.8ex] & \multicolumn{1}{c}{(1)} & \multicolumn{1}{c}{(2)} & \multicolumn{1}{c}{(3)}\\ 
\hline \\[-1.8ex] 
 Home income (log) & -0.106^{***} & -0.031^{***} & -0.075^{***} \\ 
  & (0.004) & (0.003) & (0.006) \\ 
  & & & \\ 
 Home population (log) & -0.057^{***} & 0.013^{***} & 0.061^{***} \\ 
  & (0.004) & (0.003) & (0.006) \\ 
  & & & \\ 
 Degree & -0.001^{***} & -0.001^{***} & 0.002^{***} \\ 
  & (0.0001) & (0.00004) & (0.0001) \\ 
  & & & \\ 
 Constant & 1.042^{***} & 0.242^{***} & 0.638^{***} \\ 
  & (0.024) & (0.017) & (0.033) \\ 
  & & & \\ 
\hline \\[-1.8ex] 
Metro FE & \multicolumn{1}{c}{Yes} & \multicolumn{1}{c}{Yes} & \multicolumn{1}{c}{Yes} \\ 
\hline \\[-1.8ex] 
Observations & \multicolumn{1}{c}{86,177} & \multicolumn{1}{c}{74,900} & \multicolumn{1}{c}{74,900} \\ 
R$^{2}$ & \multicolumn{1}{c}{0.055} & \multicolumn{1}{c}{0.039} & \multicolumn{1}{c}{0.027} \\ 
Adjusted R$^{2}$ & \multicolumn{1}{c}{0.054} & \multicolumn{1}{c}{0.038} & \multicolumn{1}{c}{0.026} \\ 
\hline 
\hline \\[-1.8ex] 
\textit{Note:}  & \multicolumn{3}{r}{$^{*}$p$<$0.1; $^{**}$p$<$0.05; $^{***}$p$<$0.01} \\ 
\end{tabular} 
\justify
\footnotesize{\textbf{Table 2} Linear regression models with metropolitan area fixed effects. The number of observations is smaller in case of clustering and tie support, because their calculation requires at least 2 ties within 10~km distance}
\end{table}

Besides their structural features, we also investigate the homogeneity of the individual ego networks in terms of income level. To this end, we decompose individual ego-networks analyzed above to income-based sub-networks that contain friends who live in either above or below median income level neighborhoods. This enables us to compare the structural characteristics of social networks across income groups. 

Figure~5 shows income segregation in all three approaches of social capital measurement. In Figure 5a, we find that residents of higher income neighborhoods have on average almost 60\% of their ties to users living also in higher income neighborhoods. Income homophily has an even greater role in the case of residents of lower-income neighborhoods: 70\% of their friends live in lower-income neighborhoods.  Similar patterns are found in terms of the clustering coefficient, and the share of supported ties. Users living in relatively poor census tracts tend to have higher closure in their networks among their friends who also come from lower income neighborhoods. The tie support measure further corroborates this observation, most of the ties users living in less affluent neighborhoods have to likewise people are supported relationships. Though homophily is also apparent in both of these measures among those coming from higher-income neighborhoods, its extent is always superior in case they are from lower-income neighborhoods. Taken together, income segregation is prevalent in social networks that span across neighborhoods in cities and can harm residents of lower-income neighborhoods who are typically connected to and accumulate social capital with lower-income residents. 

\begin{figure}[hbt!]
\centering
\includegraphics[width=0.9\linewidth]{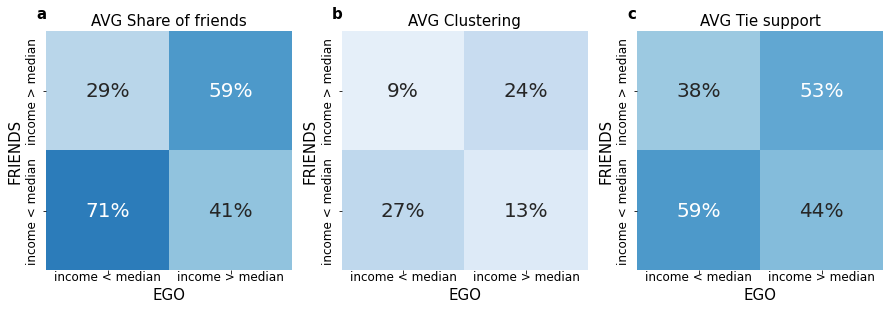}
\justify
\footnotesize{\textbf{Figure 5} Homophily among the social networks of users within 10~km distance.} 
\label{fig:fig5}
\end{figure}

\section{Discussion}

Social capital is key for prosperity in life. This work contributes to the literature on social capital by illustrating that social network features of individuals measuring some aspects of social capital are spatially concentrated in urban areas. This adds to the discussion in several points. 

First, we show that connections of individuals concentrate within only a few~kilometers around the place of residence and these connections tend to be closed and supported.
We find that spatial concentration of social capital is stronger for people living in lower-income neighborhoods, which is a general notion across the top 50 metropolitan areas of the US. The spatial concentration of social ties and social capital inside metropolitan areas also shows that like in (\cite{Bailey2020}), edge formation probability is distance dependent within cities as well as on a larger scale (\cite{liben2005geographic,Lengyel2015}) making spatial constraints important in the formation of the social fabric of cities. The physical proximity of home neighborhoods are clearly crucial places for social capital accumulation, especially for the unprivileged. Therefore, social housing and redevelopment policies have to focus on changes in access to resources and norms embedded in people's social networks when relocating individuals. 

Second, we adopt the novel individual social capital related network measure of (\cite{Jackson2019}) to a spatial social network. Results suggest that similarly to the clustering of social ties, supported relationships are also highly concentrated inside 10~kilometers. The negative coefficient of income predicting the share of supported ties in Table~3 is able to support the relationship between lower income and spatially concentrated social capital similarly to the clustering coefficient. 

Third, the pattern that the social capital of people living in lower income areas is more concentrated in urban space connects to several other works on capturing societal challenges through online social media data (\cite{Bailey2020, Wang2018, Morales2019, Dong2019, Bokanyietal2021}). Multiple studies address the problem that online social networks and interactions mirror offline segregation and inequality within metropolitan areas (\cite{Tothetal2021inequality}). Our work is able to show the micro-foundations of these segregation patterns at the individual ego-network level. It is important to note here that our measures are not directly dependent on the structure of the whole network of the metropolitan area, neither are they calculated from area-based metrics. Therefore, this large-scale ego-network approach is unique, especially at its scale, and we hope that future research will be able to make use of the anonymized network metrics published alongside this paper.

Our approach has some limitations. The definition of a social connection in this paper is based on mutual followership on the Twitter online social networking platform. While mutual followership signifies a mutual attention, due to biases both in the user-base and the sampling of the data, these relationships might not fully capture offline social relationships. Also, the definition itself might be unable to include other types of relationships that could be significant in the measurement of social capital. Information is  not  available on the nature of social ties  i.e. whether two mutual followers are co-workers, friends from  school or college. Future research might look into the concentration of social capital by investigating labeled and weighted social networks.

While the number of users  within metropolitan areas included in our analysis is proportional to the population, the user samples can be unrepresentative of the whole population regarding age, gender, income, and ethnicity. Within US Twitter users, African Americans are overrepresented (\cite{Hargittai2011}) while other ethnicities might be underrepresented (\cite{Mislove2011,Malik2015}). The users in our sample are predominantly young and well-educated (\cite{Webster2010, Sloan2015}). Therefore, we might not be able to generalize our findings to the whole population of these metropolitan areas. (\cite{Pfeffer2018}) suggests that the free 1\% sample from Twitter Streaming API that was used for the initial data collection is prone to errors because of bot activity. By imposing strict count limits, spatio-temporal constraints and mutual followership for ties, we tried to make our sample less distorted in this aspect. Also, (\cite{Morstatter2013}) and (\cite{Morstatter2014}) confirm that tweets filtered to containing GPS coordinates are retrieved to almost 90\% of the time when compared to the full dataset.

Our study uses neighborhood-level information from the American Community Survey to infer the socio-economic status of users by matching their identified home location to census tracts. The survey was conducted in 2012, while the data from Twitter was collected during 2012-2013, which again allows for some degree of bias. The socio-economic characteristics of the census tracts may be subjects to change over time and the individuals can also move to other neighborhoods. Finally, we did not consider deviations of income from the mean. 

This study does not follow changes in the network structure or socio-economic indicators of people over time. A promising extension of this line of research would be to track how social capital changes when moving to different areas of a city. Also, adding context to social ties, e.g. coworker ties, school ties might be possible from different data source, such as social networks inferred from register data. Tie context might also be derived based on natural language processing of the messages sent on the Twitter platform. 

\section*{Acknowledgements}

Eszter Bok\'anyi was supported by the ÚNKP-20-4 New National Excellence Program of the Ministry for Innovation and Technology from the source of the National Research, Development and Innovation Fund of Hungary. We thank for the usage of ELKH Cloud (\url{https://science-cloud.hu/}) that significantly helped us achieving the results published in this paper.

\section*{Conflict of interest}

The authors declare no conflict of interests.

\newpage
\printbibliography
\newpage

\clearpage

\section*{Supplementary Material}

\setcounter{equation}{0}

\subsection*{S1 - Distribution of users across the US}

\justify
Map of the selected 50 metropolitan areas in the United States and the number of users we follow in each metropolitan areas. Every user in the data have at least 10 friends with identified home location in the respective metropolitan area.

\begin{figure}[htb]
\centering
\includegraphics[width=0.9\linewidth]{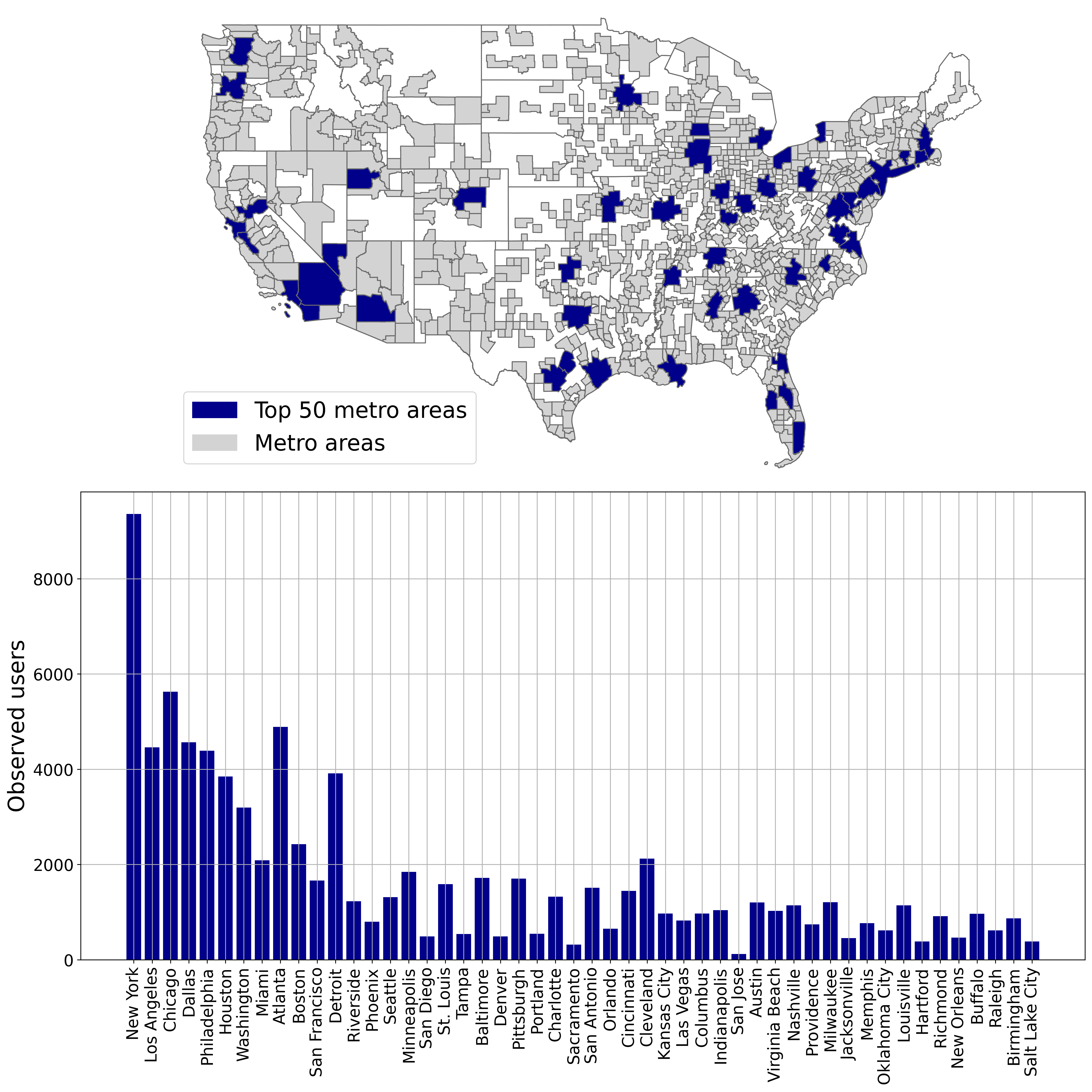}
\justify
\footnotesize{\textbf{Figure~S1} Top 50 metropolitan areas of the US and number of users in each metropolitan area.} 
\label{fig:fig1_appendix}
\end{figure}

\clearpage

\subsection*{S2 - Robustness checks}

\justify
Additional regression models that supplement the models in Table 2 of the main text. 

\begin{table}[hbt!] \centering 
\small 
\begin{tabular}{@{\extracolsep{10pt}}lD{.}{.}{-3} D{.}{.}{-3} D{.}{.}{-3} } 
\\[-1.8ex]\hline 
\hline \\[-1.8ex] 
 & \multicolumn{3}{c}{Share of friends} \\ 
\cline{2-4} 
\\[-1.8ex] & \multicolumn{1}{c}{In 10 km} & \multicolumn{1}{c}{In 5 km} & \multicolumn{1}{c}{In 1 km} \\ 
\\[-1.8ex] & \multicolumn{1}{c}{(1)} & \multicolumn{1}{c}{(2)} & \multicolumn{1}{c}{(3)}\\ 
\hline \\[-1.8ex] 
 Home income (log) & -0.106^{***} & -0.092^{***} & -0.007^{***} \\ 
  & (0.004) & (0.003) & (0.001) \\ 
  & & & \\ 
 Home population (log) & -0.057^{***} & -0.043^{***} & -0.021^{***} \\ 
  & (0.004) & (0.003) & (0.001) \\ 
  & & & \\ 
 Degree & -0.001^{***} & -0.001^{***} & -0.0002^{***} \\ 
  & (0.0001) & (0.00005) & (0.00002) \\ 
  & & & \\ 
 Constant & 1.042^{***} & 0.750^{***} & 0.137^{***} \\ 
  & (0.024) & (0.019) & (0.007) \\ 
  & & & \\ 
\hline \\[-1.8ex] 
Metro FE & \multicolumn{1}{c}{Yes} & \multicolumn{1}{c}{Yes} & \multicolumn{1}{c}{Yes} \\ 
\hline \\[-1.8ex] 
Observations & \multicolumn{1}{c}{86,177} & \multicolumn{1}{c}{86,177} & \multicolumn{1}{c}{86,177} \\ 
R$^{2}$ & \multicolumn{1}{c}{0.055} & \multicolumn{1}{c}{0.061} & \multicolumn{1}{c}{0.027} \\ 
Adjusted R$^{2}$ & \multicolumn{1}{c}{0.054} & \multicolumn{1}{c}{0.060} & \multicolumn{1}{c}{0.026} \\ 
\hline 
\hline \\[-1.8ex] 
\textit{Note:}  & \multicolumn{3}{r}{$^{*}$p$<$0.1; $^{**}$p$<$0.05; $^{***}$p$<$0.01} \\ 
\end{tabular} 
\justify
\footnotesize{\textbf{Table~S1} Linear regression models with metropolitan area fixed effects. Dependent variable: Share of friends within 10, 5 and 1 km respectively. Independent variables: natural logarithm of average income level and population in users' home neighborhoods, overall degree of users.
} 
\end{table}

\begin{table}[hbt!] \centering 
\small 
\begin{tabular}{@{\extracolsep{10pt}}lD{.}{.}{-3} D{.}{.}{-3} D{.}{.}{-3} } 
\\[-1.8ex]\hline 
\hline \\[-1.8ex] 
 & \multicolumn{3}{c}{Clustering coefficient} \\ 
\cline{2-4} 
\\[-1.8ex] & \multicolumn{1}{c}{In 10 km} & \multicolumn{1}{c}{In 5 km} & \multicolumn{1}{c}{In 1 km} \\ 
\\[-1.8ex] & \multicolumn{1}{c}{(1)} & \multicolumn{1}{c}{(2)} & \multicolumn{1}{c}{(3)}\\ 
\hline \\[-1.8ex] 
 Home income (log) & -0.031^{***} & -0.035^{***} & 0.006^{**} \\ 
  & (0.003) & (0.003) & (0.003) \\ 
  & & & \\ 
 Home population (log) & 0.013^{***} & -0.001 & -0.019^{***} \\ 
  & (0.003) & (0.003) & (0.003) \\ 
  & & & \\ 
 Degree & -0.001^{***} & -0.001^{***} & -0.0005^{***} \\ 
  & (0.00004) & (0.00003) & (0.00003) \\ 
  & & & \\ 
 Constant & 0.242^{***} & 0.247^{***} & 0.080^{***} \\ 
  & (0.017) & (0.014) & (0.016) \\ 
  & & & \\ 
\hline \\[-1.8ex] 
Metro FE & \multicolumn{1}{c}{Yes} & \multicolumn{1}{c}{Yes} & \multicolumn{1}{c}{Yes} \\ 
\hline \\[-1.8ex] 
Observations & \multicolumn{1}{c}{74,900} & \multicolumn{1}{c}{59,993} & \multicolumn{1}{c}{14,109} \\ 
R$^{2}$ & \multicolumn{1}{c}{0.039} & \multicolumn{1}{c}{0.045} & \multicolumn{1}{c}{0.075} \\ 
Adjusted R$^{2}$ & \multicolumn{1}{c}{0.038} & \multicolumn{1}{c}{0.044} & \multicolumn{1}{c}{0.071} \\ 
\hline 
\hline \\[-1.8ex] 
\textit{Note:}  & \multicolumn{3}{r}{$^{*}$p$<$0.1; $^{**}$p$<$0.05; $^{***}$p$<$0.01} \\ 
\end{tabular} 
\justify
\footnotesize{\textbf{Table~S2} Linear regression models with metropolitan area fixed effects. Dependent variable: Local clustering coefficient within 10, 5 and 1 km respectively. Independent variables: natural logarithm of average income level and population in users' home neighborhoods, overall degree of users.
} 
\end{table}

\begin{table}[hbt!] \centering 
\small 
\begin{tabular}{@{\extracolsep{10pt}}lD{.}{.}{-3} D{.}{.}{-3} D{.}{.}{-3} } 
\\[-1.8ex]\hline 
\hline \\[-1.8ex] 
 & \multicolumn{3}{c}{Tie support} \\ 
\cline{2-4} 
\\[-1.8ex] & \multicolumn{1}{c}{In 10 km} & \multicolumn{1}{c}{In 5 km} & \multicolumn{1}{c}{In 1 km} \\ 
\\[-1.8ex] & \multicolumn{1}{c}{(1)} & \multicolumn{1}{c}{(2)} & \multicolumn{1}{c}{(3)}\\ 
\hline \\[-1.8ex] 
 Home income (log) & -0.075^{***} & -0.094^{***} & -0.066^{***} \\ 
  & (0.006) & (0.007) & (0.015) \\ 
  & & & \\ 
 Home population (log) & 0.061^{***} & 0.059^{***} & 0.005 \\ 
  & (0.006) & (0.007) & (0.017) \\ 
  & & & \\ 
 Degree & 0.002^{***} & 0.002^{***} & -0.0003 \\ 
  & (0.0001) & (0.0001) & (0.0002) \\ 
  & & & \\ 
 Constant & 0.638^{***} & 0.672^{***} & 0.704^{***} \\ 
  & (0.033) & (0.042) & (0.097) \\ 
  & & & \\ 
\hline \\[-1.8ex] 
Metro FE & \multicolumn{1}{c}{Yes} & \multicolumn{1}{c}{Yes} & \multicolumn{1}{c}{Yes} \\ 
\hline \\[-1.8ex] 
Observations & \multicolumn{1}{c}{74,900} & \multicolumn{1}{c}{59,993} & \multicolumn{1}{c}{14,109} \\ 
R$^{2}$ & \multicolumn{1}{c}{0.027} & \multicolumn{1}{c}{0.016} & \multicolumn{1}{c}{0.015} \\ 
Adjusted R$^{2}$ & \multicolumn{1}{c}{0.026} & \multicolumn{1}{c}{0.015} & \multicolumn{1}{c}{0.011} \\ 
\hline 
\hline \\[-1.8ex] 
\textit{Note:}  & \multicolumn{3}{r}{$^{*}$p$<$0.1; $^{**}$p$<$0.05; $^{***}$p$<$0.01} \\ 
\end{tabular} 
\justify
\footnotesize{\textbf{Table~S3} Linear regression models with metropolitan area fixed effects. Dependent variable: Share of supported ties within 10, 5 and 1 km respectively. Independent variables: natural logarithm of average income level and population in users' home neighborhoods, overall degree of users.
} 
\end{table} 

\clearpage
\subsection*{S3 - Linear models behind Figure 4}

Figure 4 illustrates how the dummy variables (0/1) on home income $>$ median income in the respective metro area ($HI_{I>m}$) correlates with share of friends ($SoF_d$=10)), local clustering coefficient ($Clust_{d=10}$) and supported ties ($Supp_{d=10}$) inside 10~kilometer ($d=10$) from the home location of individual users. We estimate the relationship between home income and the different network variables by controlling for the log transferred population ($Pop$) of the home census tract. Figure 4 visualizes the $\beta_1$ coefficients of the linear regression models below:

\begin{equation}
SoF_{d=10} = \beta_{0} +\beta_{1}HI_{I>m} + \beta_{2}Pop
\end{equation}

\begin{equation}
Clust_{d=10} = \beta_{0} +\beta_{1}HI_{I>m} + \beta_{2}Pop
\end{equation}

\begin{equation}
Supp_{d=10} = \beta_{0} +\beta_{1}HI_{I>m} + \beta_{2}Pop
\end{equation}

\end{document}